\documentclass[twocolumn,showpacs,preprintnumbers,amsmath,amssymb]{revtex4}

\usepackage{graphicx}
\usepackage{dcolumn}
\usepackage{bm}

\begin{document}

\preprint{APS/123-QED}

\title{Anomalous elastic softening of SmRu$_{4}$P$_{12}$ under high pressure}

\author{Peijie Sun}
\author{Yoshiki Nakanishi}%
\author{Mitsuteru Nakamura}
\author{Masahito Yoshizawa}
 \email{yoshizawa@iwate-u.ac.jp}
\affiliation{%
Graduate School of Engineering, Iwate University, Morioka 020-8551, Japan
}%

\author{Masashi Ohashi}
\author{Gendo Oomi}
\affiliation{
Department of Physics, Kyushu University, Fukuoka 810-8560, Japan 
}%

\author{Chihiro Sekine}
\author{Ichimin Shirotani}
\affiliation{
Faculty of Engineering, Muroran Institute of Technology, Muroran 050-8585, Japan
}%

\date{\today}

\begin{abstract}
The filled skutterudite compound SmRu$_4$P$_{12}$ undergoes a complex evolution from a paramagnetic metal (phase I) to a 
probable multipolar ordering insulator (phase II) at $T_{\rm MI}$\,$\sim$\,16.5 K,  then to a magnetically ordered phase 
(phase III) at $T_{\rm N}$\,$\sim$\,14 K.     
Elastic properties under hydrostatic pressures were investigated to study the nature of the ordering phases. We found that 
distinct elastic softening above  $T_{ \rm MI}$ is induced by pressure, giving evidence of quadrupole degeneracy of the 
ground state in the crystalline electric field. 
It also suggests that quadrupole moment may be one of the order parameters below $T_{\rm MI}$ under pressure.
Strangely, the largest degree of softening is found in the transverse elastic constant $C_{\rm T}$ at around 0.5-0.6 GPa, 
presumably having relevancy to the competing and very different Gr\"{u}neisen parameters $\Omega$ of $T_{\rm MI}$ and 
$T_{\rm N}$. 
Interplay between the two phase transitions is also verified by the rapid increase of $T_{ \rm MI}$ under pressure with a 
considerably large $\Omega$ of 9. Our results can be understood on the basis of the proposed octupole scenario for 
SmRu$_4$P$_{12}$. 
\end{abstract}


\maketitle

\section{Introduction}

SmRu$_{4}$P$_{12}$ is one of the more interesting members of the family of filled skutterudite $RT_4X_{12}$ compounds ($R$ 
= rare earths; $T$ = Fe, Ru, Os; $X$ = P, As, Sb) which shows a variety of physical properties including 
superconductivity,  metal-insulator (MI) transition, magnetic ordering and heavy fermions (HF) 
\cite{bauer,sanada,sekine1}. The interest in SmRu$_{4}$P$_{12}$ is due to its metal-insulator (MI) transition at $T_{\rm 
MI}$\,$\sim$\,16.5 K, a subsequent antiferromagnetic (AFM) transition at $T_{\rm N}$\,$\sim$\,14 K \cite{sekine1},  and a 
strange $H$-$T$ magnetic phase diagram \cite{matsuhira1, sekine2}. $T_{\rm N}$ is obscure at lower magnetic fields; 
however, it is distinctly visible in several measurements like thermal expansion \cite{sekine4} even in a zero magnetic 
field. 
With increasing magnetic field, $T_{\rm MI}$ increases while $T_{\rm N}$ decreases, resembling CeB$_6$, which has an 
antiferro-quadrupolar (AFQ) ordering and a subsequent magnetic ordering \cite{effan}. For this reason, SmRu$_{4}$P$_{12}$ 
was initially thought to be an AFQ system; however, increasing doubt is being thrown on this view, stimulating interest   
in the order parameters (OPs).  
SmRu$_{4}$P$_{12}$ crystallizes in a cubic structure with space group Im\={3} (T$_{\rm h}^5$), like other filled 
skutterudite compounds. Magnetic measurements show that Sm is trivalent with total angular momentum $J$ = 5/2.   
Specific heat measurements suggest a crystalline electric field (CEF) scheme consisting of a $\Gamma _{67}$ ground state 
quartet and an excited doublet $\Gamma _{5}$ at about 60 K in the T$_{\rm h}$ symmetry \cite{matsuhira1,takegahara}. This 
scheme is plausible for understanding other measurements including the elastic constant \cite{yoshizawa1,yoshizawa3}.  
Noticeably, the $\Gamma _{67}$ quartet with both orbital and magnetic degeneracy, is a key point for understanding the 
various properties of SmRu$_{4}$P$_{12}$.  Hydrostatic pressures of about 1\,GPa in this work are assumed to have no 
substantial effects on the CEF scheme. 

The AFQ scenario for SmRu$_{4}$P$_{12}$ is under question as to several aspects. First, almost no elastic softening is 
observed above $T_{\rm MI}$ \cite{yoshizawa1} and the elastic anomaly at $T_{\rm MI}$ is not very large, unlike a typical 
AFQ ordering such as in CeB$_6$ \cite{nakamura} and DyB$_2$C$_2$ \cite{nemoto}. Second, by application of magnetic fields, 
the specific heat anomaly of AFQ ordering in CeB$_6$ is enhanced, while that at $T_{\rm MI}$ in SmRu$_{4}$P$_{12}$ changes 
only slighy; meanwhile, the anomaly at $T_{\rm N}$ in CeB$_6$ is weakened, whereas it is enhanced in SmRu$_{4}$P$_{12}$ 
\cite{effan,matsuhira2}. 
Recently, the possibility of octupole ordering was proposed for the MI transition \cite{yoshizawa2}. This proposition is 
based on the appearance of elastic softening in phase II toward $T_{\rm N}$ and the indistinctness of the subsequent 
magnetic ordering at $T_{\rm N}$, because these facts suggest a new coupling of $\Psi M\epsilon$ ($\Psi$: probable 
octupole OP in phase II; $M$: dipole OP in phase III; $\epsilon$: elastic strain induced by ultrasound) and indicate 
breakdown of the time reversal symmetry (TRS) in phase II. Breakdown of TRS is confirmed by muon spin relaxation ($\mu$SR) 
\cite{hachitani}, nuclear magnetic resonance (NMR) \cite{masaki} and Sm nuclear resonant scattering \cite{tsutsui} 
measurements. An AFQ ordering is nonmagnetic and holds TRS, so is ruled out as the primary OP in phase II. On the other 
hand, an octupole ordering is magnetic and breaks TRS, so is possible in phase II. Microscopic measurements also suggest 
appearance of magnetic dipole moment in phase II even in a zero magnetic field, in addition to an unknown multipole 
ordering \cite{masaki,tsutsui}. This is very different to the strong candidate for octupole ordering material, NpO$_2$ 
\cite{paixao}, which has only one phase transition.  

An iso-structural compound, PrRu$_{4}$P$_{12}$, also shows MI transition ($T_{\rm MI}$\,$\simeq$\,63 K) \cite{sekine3}. 
Its MI transition is accompanied by no magnetic anomaly, whereas a distinct magnetic anomaly is observed in 
SmRu$_{4}$P$_{12}$. Moreover, a structural distortion is observed in PrRu$_{4}$P$_{12}$ \cite{lee} but is absent in 
SmRu$_{4}$P$_{12}$. 
These facts indicate a magnetic origin for the MI transition in SmRu$_{4}$P$_{12}$, unlike PrRu$_{4}$P$_{12}$. In fact, a 
charge density wave (CDW) due to Fermi surface nesting is plausible for interpreting the MI transition in 
PrRu$_{4}$P$_{12}$ \cite{harima}. Recent X-ray and polarized neutron diffraction experiments have shown that the Pr-ion 
site splits into two nonequivalent crystallographic sites and suggest two different CEF schemes below $T_{\rm MI}$ 
\cite{iwasa}, where the importance of the $p$-$f$ hybridization in the MI transition is also emphasized.   

Ultrasonic measurement under hydrostatic pressures is rarely performed, partly due to its difficulty. We succeeded, 
however, in obtaining stable ultrasonic echoes by selecting a suitable sample of sufficient length.
By performing such measurements, we intend to gain an insight into the following peculiarities in SmRu$_{4}$P$_{12}$. (1) 
The $\Gamma _{67}$ quartet ground state degenerates with respect to $\Gamma _3$ and $\Gamma _5$-type quadrupole moments; 
however, no distinct elastic softening is observed at ambient pressure. Application of hydrostatic pressures is expected 
to induce stronger quadrupole-strain and inter-site quadrupole interaction, which then affects elastic properties. 
Quadrupolar ordering, if visible under pressure, is helpful for understanding the possible OPs below $T_{\rm MI}$ and the 
peculiar $H$-$T$ phase diagram.    
(2) We intend to build a pressure-temperature ($P$-$T$) phase diagram from which to gain knowledge of the OPs of phase II 
and III.  (3) One more point of interest is the very different Gr\"{u}neisen parameters $\Omega$ of $T_{\rm MI}$ and 
$T_{\rm N}$, the latter being much larger than the former \cite{yoshizawa5}. Widely different Gr\"{u}neisen parameters 
suggest different pressure dependencies of the two phase transitions, and prompted us to explore the physical properties 
under hydrostatic pressures related to this phenomenon.  

\section{Experiments}
A cylindrical polycrystalline SmRu$_4$P$_{12}$ sample 4.1\,mm in length and 3.4\,mm in diameter was employed. It is the 
same one as used for ultrasonic measurements in magnetic fields \cite{yoshizawa1} and was prepared at high temperatures  
and pressures using a wedge-type cubic anvil high-pressure apparatus. Actually, a single crystal is preferred here, but 
the available single crystals currently are very small, with dimensions of only 1$\times$1$\times$1 mm, preventing us from 
obtaining stable ultrasonic echoes under high pressures. 

The relative change of sound velocity $\upsilon$ was measured by means of a phase-comparison technique. This technique 
measures the relative  $\Delta \upsilon/\upsilon$ by detecting the relative frequency change $\Delta f/f$ while keeping 
the phase shift at zero and neglecting any changes in sample length $\Delta l/l$ (to be explained later). Elastic constant 
$C$ is related to sound velocity $\upsilon$ by the equation $C$\,=\,$\rho \upsilon ^2$, where the density $\rho$ of 
SmRu$_{4}$P$_{12}$ is 5.921g$\cdot$cm$^{-3}$.  LiNbO$_3$ piezoelectric plate with fundamental resonance frequency of 
$5$-$20$ MHz was used as the ultrasonic transducer to both generate and detect the ultrasound. The only experimentally 
accessible ultrasound modes for a polycrystalline sample, i.\,e., longitudinal $C_{\rm L}$  propagating parallel to 
polarization and transverse $C_{\rm T}$ propagating perpendicular to polarization, were measured as a function of 
temperature under different hydrostatic pressures.  
The absolute values of $C_{\rm L}$, $C_{\rm T}$ and bulk modulus $C_{\rm B}$ of SmRu$_4$P$_{12}$ at 100 K are 153, 24.6 
and 120 GPa, respectively \cite{yoshizawa1}. 

Hydrostatic pressure up to 1.15 GPa was generated using a Cu-Be piston-cylinder pressure cell and a Teflon capsule inside 
it. The cell has two layers, consisting of an outer cylinder and an inner one placed inside. The inner diameter and the 
length of the inner cylinder are 8\,mm and 40\,mm, respectively. The sample with two ultrasonic transducers bonded to it, 
together with a tin manometer, is placed in the Teflon capsule filled with pressure medium (a 1:1 mixture of Fluorinert FC 
70 and FC 77). The measurements were carried out down to the temperature of liquid helium, 4.2 K, or down to  $\sim$1.5 K 
by pumping. The effective hydrostatic pressure inside the Teflon capsule was calibrated by measuring the superconducting 
transition temperature of the tin manometer.  

\section{Experimental results}
\begin{figure}[htp]
\includegraphics[width=0.65\linewidth]{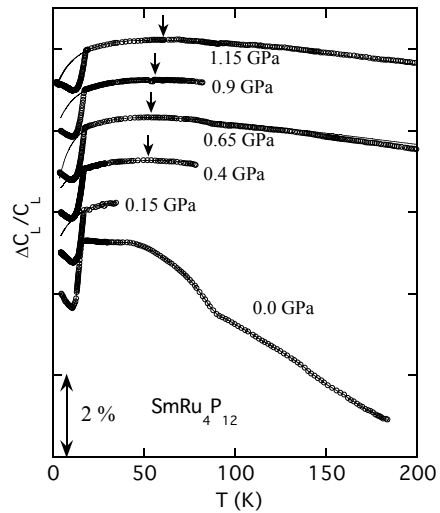}
\caption{Temperature dependence of the longitudinal elastic constant $C_{\rm L}$ under various pressures. The solid lines 
are calculated curves by Eq.~(\ref{eq:2}). The arrows indicate $T_{\rm soft}$ below which elastic softening appears.
\label{CLall.eps}}
\end{figure}
Fig.~\ref{CLall.eps} shows the temperature dependence of the longitudinal mode $\Delta C_{\rm L}/C_{\rm L}$ under various 
pressures. The solid lines are fitted curves that will be explained later.  The $C_{\rm L}$ at ambient pressure increases 
on cooling down to $T_{\rm MI}$ without elastic softening. Application of hydrostatic pressures induces a clear softening 
above $T_{\rm MI}$,  but its pressure dependence is very weak. This may be partially due to the relative hardness of  
$C_{\rm L}$ that includes bulk modulus $C_{\rm B}$. The temperature $T_{\rm soft}$, below which softening appears, shows a 
slight increase with increasing pressure.    

\begin{figure}[htp]
\includegraphics[width=0.65\linewidth]{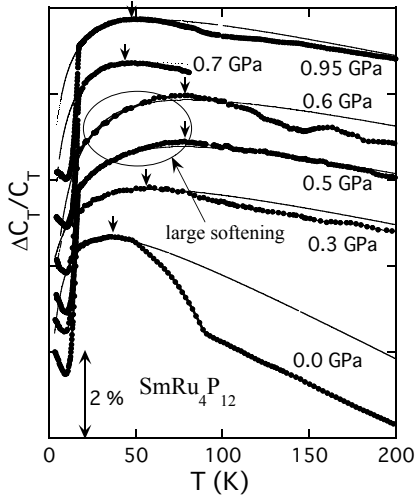}
\caption{Temperature dependence of the transverse elastic constant $C_{\rm T}$ under various pressures. The solid lines 
and arrows denote the same as in Fig.~\ref{CLall.eps}.
\label{CTall.eps}}
\end{figure}

Fig.~\ref{CTall.eps} shows the temperature dependence of the transverse mode  $\Delta C_{\rm T}/C_{\rm T}$ at various 
pressures, together with the fitted curves. Ambient pressure evidences a weak elastic softening above $T_{\rm MI}$ up to 
35 K. With increasing pressure up to 0.6 GPa, elastic softening becomes increasingly marked. Furthermore, $T_{\rm soft}$ 
increases from 35 K for 0 GPa to about 80 K for 0.6 GPa. Strangely, upon further increasing the pressure by 0.7 GPa, the 
elastic softening fades again, while at the same time, $T_{\rm soft}$ also decreases. This strange pressure dependence 
presumably reflects the complexity of the phase diagram with two competing phase transitions. A kink structure is observed 
at around 90 K in both $C_{\rm T}$ and $C_{\rm L}$ at ambient pressure, reminiscent of the off-center motion of the $R$ 
ions in the pnictogen cage, namely ``rattling'' as in PrOs$_4$Sb$_{12}$ \cite{goto}. However, here it seems not due to an 
intrinsic origin because it is absent in single crystals. The kink becomes very weak or undetectable as soon as even a 
relatively weak pressure is applied; this is also not expected for a rattling motion. 

For both $C_{\rm L}$ and $C_{\rm T}$, a distinct slope change in the background $C^0$, is noticeable i.\,e., the lattice 
contribution to elastic constant, before and after applying pressure. As for the reasons, 
the influence of pressure-induced sample length variation $\Delta l/l$, which alters the relative ultrasound velocity as 
follows, is possible. 
\begin{eqnarray}
\frac{\Delta \upsilon}{\upsilon} = \frac{\Delta f}{f} - \frac{\Delta \varphi}{\varphi} + \frac{\Delta l}{l}
\label{eq:0}.
\end{eqnarray}
Eq.~(\ref{eq:0}) shows that in the phase-comparison technique, relative sound velocity $\Delta \upsilon/\upsilon$ is 
composed of three parts, e.\,g., the relative frequency $\Delta f/f$, relative signal phase $\Delta \varphi/\varphi$ and 
sample length $\Delta l/l$. $\Delta \varphi/\varphi$ is kept at zero by a feedback loop that compares the phase shift 
between the ultrasonic signal transported by the sample and the reference signal. $\Delta l/l$ is usually negligibly 
small. Therefore, this equation also explains why ultrasonic velocity $\upsilon$ is measured as a shift in frequency. 
If $\Delta l/l$ induced by pressure accounts for the change in $C^0$, the linear thermal expansion coefficient 
$\alpha$($T$) is anticipated to be a change over 5 $\times$ $10^{-5}$ K$^{-1}$ under a pressure of about 0.3 GPa, in 
comparison with that at ambient pressure. This is at least an order of magnitude greater than the $\alpha$($T$) at ambient 
pressure for SmRu$_4$P$_{\rm 12}$ \cite{sekine4}, making it impossible for $\Delta l/l$  to account for the change in 
$C^0$.  Although this peculiarity is not well understood, it does not affect our analysis of the elastic softening at low 
temperatures.    

The temperature where a steep elastic drop occurs is known as the MI transition temperature $T_{\rm MI}$ at ambient 
pressure. Similarly, all such anomalies observed under pressure are ascribed to MI transition. Although significant 
elastic softening is induced by hydrostatic pressure, the shape of the abrupt anomaly at $T_{\rm MI}$ shows no substantial 
change. In fact, electrical resistance under high pressures remains characteristic of MI transition up to at least 1.2 
GPa, while metallic behavior appears at $P$\,$>$\,3.5\,GPa \cite{miyake}. $T_{\rm MI}$ increases on applying pressure, in 
agreement with the observations in electrical resistance. This issue will be discussed later in terms of Gr\"{u}neisen 
parameter $\Omega$. Unfortunately, we can not detect an anomaly at $T_{\rm N}$ under pressure, as was successfully done in 
magnetic fields \cite{yoshizawa1}. 

\section{Discussion}
\subsection{Pressure-induced elastic softening}
Symmetrized strains induced by ultrasound perturb CEF state via quadrupole-strain interaction $g_{\rm \Gamma} O_{\Gamma} 
\epsilon _{\Gamma}$. Taking the inter-site quadrupole interaction $g_{\Gamma}^{\prime}O_{\Gamma}\langle O_{\Gamma}\rangle$ 
into account, elastic constant $C_\Gamma$ can be formulated as follows by the second derivative of Landau free energy. 
\begin{eqnarray}
C_\Gamma(T) &=& C_\Gamma^0(T)-\frac{Ng_\Gamma^2\chi_\Gamma(T)}{1-g_\Gamma^\prime\chi_\Gamma(T)}
\label{eq:1}.
\end{eqnarray}
Here $C_\Gamma^0(T)$ is background elastic constant, $N$ the number of $R$ ions per unit volume, $\chi_\Gamma(T)$ the 
corresponding strain susceptibility calculated using the CEF scheme, and $g_\Gamma$ and $g_\Gamma^\prime$ are the coupling 
constants of the quadrupole-strain and quadrupole-quadrupole, respectively.  $C_\Gamma(T)$ measures the diagonal (Curie 
term) and non-diagonal (Van-Vleck term) quadrupolar matrix elements, analogous to the magnetic susceptibility that 
measures magnetic dipole matrix elements \cite{luthi}.
Analyzing the experimental results by using $\chi_ {\rm \Gamma}(T)$, one experimentally obtains $|g_{\rm \Gamma}|$ and 
$g_{\rm \Gamma}^\prime$. 

At ambient pressure, no elastic softening was observed above $T_{\rm MI}$ for $C_{\rm L}$ and only very slight softening 
for $C_{\rm T}$. Generally, this indicates a very weak quadrupole-strain coupling $g_{\Gamma}$ for the $\Gamma _{\rm 67}$ 
quartet ground state. This is believed to be a signature of  complex multipolar ordering in phase II.  On the other hand, 
a high degree of softening is observed under hydrostatic pressures, especially for $C_{\rm T}$, giving evidence of the 
degeneracy  of the CEF ground state with respect to quadrupole moment.
Hydrostatic pressure induces elastic strain $\epsilon _{B}$\,=\,$\epsilon _{xx} + \epsilon _{yy} +\epsilon _{zz}$ that 
corresponds to bulk modulus $C_{\rm B}$. $\epsilon _{B}$ has no direct couple with symmetry quadrupole moment 
$O_{\Gamma}$, thus is not a direct reason for the enhanced elastic softening. 

A crucial point in the physics of the enhanced elastic softening is which of the two types of inherent quadrupole moments 
($\Gamma _3$ or $\Gamma _5$) is responsible. The answer to this question may give direct insight into the OP symmetry of 
phase II; unfortunately, it can not be clarified based on the present results on a polycrystalline sample. 
In contrast to the symmetrized transverse modes in a single crystal like $(C_{\rm 11}-C_{\rm 12})/2$ and $C_{\rm 44}$ that 
respectively corresponds to $\Gamma _{3}$ and $\Gamma _{5}$-type quadrupole moment, $C_{\rm T}$ and $C_{\rm L}$ in a 
polycrystalline sample reflect an average of different elastic modes. The former consists of $C_{\rm 44}$ and 
$(C_{11}-C_{12})/2$; the latter all the available modes including $C_{\rm 11}$ and bulk modulus $C_{\rm B}$. 
For the same reason, Eq.~(\ref{eq:1}) cannot be directly used to analyze elastic softening here. 
Nevertheless, as long as the $\chi_ {\rm \Gamma}(T)$ is dominated by the Curie term as predicted in the present system, 
Eq.~(\ref{eq:1}) can be reformulated as follows. 
\begin{eqnarray}
C_\Gamma(T) &=& C_\Gamma^0(T) \Big(\frac{T-T_C^0}{T-T_Q}\Big)
\label{eq:2}.
\end{eqnarray}
According to Eq.~(\ref{eq:2}), the quadrupole-strain coupling and inter-site quadrupole interaction can be evaluated as an 
average. Furthermore, transforming Eq.~(\ref{eq:2}) to Eq.~(\ref{eq:3}) is helpful in actual analysis. 
\begin{eqnarray}
\frac{1}{C_\Gamma^0(T)-C_\Gamma(T)} \approx \frac{T-T_Q}{T_C^0-T_Q}
\label{eq:3}.
\end{eqnarray} 
Here $T_{\rm Q}$\,=\,$g_\Gamma^\prime$$|\langle\varphi _i |O_\Gamma|\varphi _i\rangle|^2$ indicates the inter-site 
quadrupole interaction. 
The difference of the two characteristic temperatures, $T_{\rm C}^0-T_{\rm Q}$\,=\,$Ng_\Gamma^2$$|\langle\varphi _i 
|O_\Gamma|\varphi _i\rangle|^2$/$C_\Gamma^0$, is usually termed the Jahn-Teller coupling energy $E_{\rm JT}$\cite{what1}, 
and is a function of quadrupole-strain coupling.  

\begin{figure}[htp]
\includegraphics[width=0.65\linewidth]{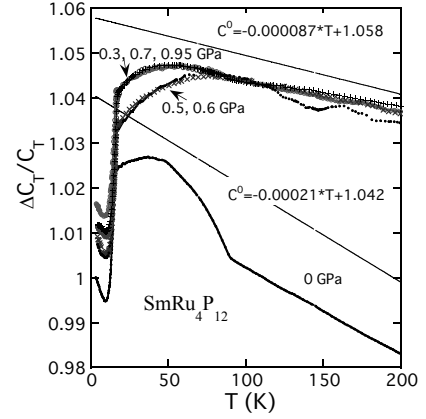}
\caption{Transverse elastic constant $C_{\rm T}(T)$ under various pressures that is superposed on each other. The two 
solid lines indicate estimated elastic background, described as $C^0 = -0.00021 \times T + 1.042$ and $C^0 = -0.000087 
\times T + 1.058$, for  0 GPa and other non-ambient pressures, respectively.
\label{CTC0.eps}}
\end{figure}
Before analyzing elastic softening, elastic background $C^0(T)$ should be carefully estimated. This process may affect the 
fitted values of $T_{\rm Q}$ and $E_{\rm JT}$ to some degree. Here we employ the commonly used $T$-linear $C^0$. 
Fig.~\ref{CTC0.eps} presents the measurement data of $C_{\rm T}$ and two estimated $C^0(T)$ lines. All the curves at 
non-ambient pressures are superposed on each other and a common $C^0(T)$ is estimated for them. Significantly, we found 
the pressure dependences of $T_{\rm Q}$ and $E_{\rm JT}$ to be almost independent of $C^0(T)$ as long as a common 
background is employed as done in this Figure. In fact, a shared slope of all $C_{\rm T}(T)$ curves under non-ambient 
pressures at temperatures over 100 K can be observed, supporting the common $C^0(T)$. This procedure can maximally reduce 
extrinsic influence in estimating the two characteristic temperatures, and is crucial for the later discussions.   

\begin{figure}[htp]
\includegraphics[width=0.65\linewidth]{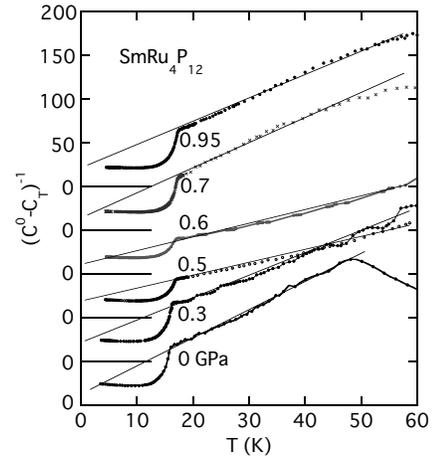}
\caption{($C^0$-$C_{\rm T}$)$^{-1}$ as a function of temperature under various pressures. The solid lines are Curie term 
fits based on Eq.~(\ref{eq:3}), which give values of $T_{\rm Q}$ and $E_{\rm JT}=T_{\rm C}^0-T_{\rm Q}$.
\label{curiefit.eps}}
\end{figure}
Fig.~\ref{curiefit.eps} shows ($C^0$-$C_{\rm T}$)$^{-1}$ as a function of temperature. The $T$-linear increase above 
$T_{\rm MI}$ conforms to the Curie term described by Eq.~(\ref{eq:3}), due to the $\Gamma _{\rm 3}$ and/or $\Gamma _{\rm 
5}$-type quadrupole moment of the $\Gamma _{\rm 67}$ ground state.
The horizontal intercepts and the slopes of the linear fits give the quadrupolar interaction $T_{\rm Q}$ and Jahn-Teller 
coupling energy $E_{\rm JT}$. For longitudinal $C_{\rm L}$ we performed the same procedure (not shown here). Note that the 
calculated curves shown in Fig.~\ref{CLall.eps},~\ref{CTall.eps} are based on the obtained $T_{\rm Q}$ and $E_{\rm JT}$ 
values, and Eq.~(\ref{eq:2}). 

\begin{figure}[htp]
\includegraphics[width=0.65\linewidth]{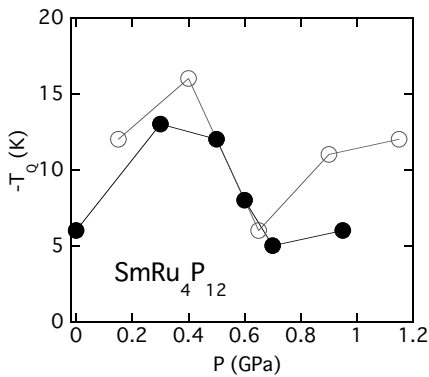}
\caption{Quadrupolar interaction $T_{\rm Q}$ as a function of pressure. Open circles denote data estimated from $C_{\rm 
L}$ and solid circles from $C_{\rm T}$. 
\label{TQ.eps}}
\end{figure}
   
\begin{figure}[htp]
\includegraphics[width=0.65\linewidth]{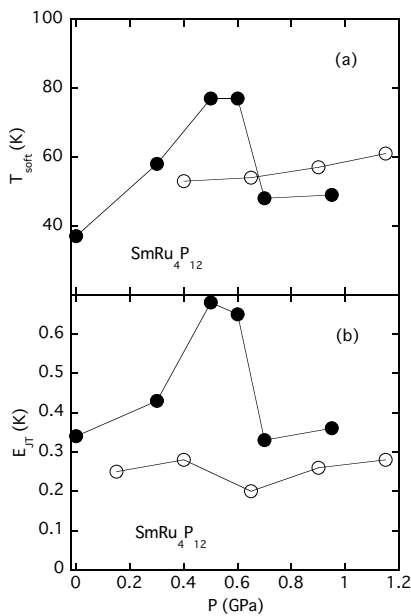}
\caption{Pressure dependences of  (a) the temperature $T_{\rm soft}$ below which elastic softening appears, and (b) the 
Jahn-Teller energy $E_{\rm JT}$ = $T_{\rm C}^{\rm 0}$-$T_{\rm Q}$. Open circles denote data estimated from $C_{\rm L}$ and 
solid circles from $C_{\rm T}$.  
\label{Tsoft.eps}}
\end{figure}
Fig.~\ref{TQ.eps} presents pressure dependence of the quadrupolar interaction, $-T_{\rm Q}$ vs. P.  
$T_{\rm Q}$s are all negative in indication of antiferro-quadrupolar interaction. $T_{\rm Q}$ is once enhanced by 
pressure; however, it is weakened again at around 0.7 GPa where elastic softening is weakened.  Shown in 
Fig.~\ref{Tsoft.eps} are $T_{\rm soft}$ and  Jahn-Teller energy $E_{\rm JT}$ as a function of pressure. Both of them 
depend on the same quadrupole-strain coupling and reflect the lattice instability. As expected, $T_{\rm soft}$ and $E_{\rm 
JT}$ show a similar pressure dependence. However, the two energies exhibit different pressure dependences for $C_{\rm T}$ 
and $C_{\rm L}$;  enhanced values around 0.5-0.6 GPa are found in transverse $C_{\rm T}$, while no distinct change is 
observed in longitudinal $C_{\rm L}$. 

Here we first discuss the pressure dependence of $T_{\rm Q}$ and $E_{\rm JT}$. Because $T_{\rm 
Q}$\,=\,$g_\Gamma^\prime$$|\langle\varphi _i |O_\Gamma|\varphi _i\rangle|^2$ and $E_{\rm 
JT}$\,=\,$Ng_\Gamma^2$$|\langle\varphi _i |O_\Gamma|\varphi _i\rangle|^2$/$C_\Gamma^0$, two possible explanations of their 
pressure dependences are available. (1) The two energies change with pressure due to changes in quadrupole moment 
$|\langle\varphi _i |O_\Gamma|\varphi _i\rangle|$. (2) Their changes are due to $g_\Gamma^\prime$ and $g_\Gamma$, 
respectively. For case (1), one expects a similar pressure dependence of $T_{\rm Q}$ and $E_{\rm JT}$. This is different 
to our observations. Moreover, a significant change of the CEF scheme and the consequent quadrupole moment are not 
anticipated under the present hydrostatic pressures. The quadrupole Kondo effect may have an influence on $|\langle\varphi 
_i |O_\Gamma|\varphi _i\rangle|$ but one would expect a screened quadrupole moment by applying pressure, not an enhanced 
one.  Thus, it seems likely that changes in $g_\Gamma^\prime$ and $g_\Gamma$ can be ascribed to the characteristic 
energies under pressure. Because $g_\Gamma^\prime$$\sim$$T_{\rm Q}$ is negative and $g_\Gamma^2$$\sim$$E_{\rm JT}$ is 
positive as factors in Eq.~(\ref{eq:1}), the rapid decline of $|T_{\rm Q}|$ and enhancement of $E_{\rm JT}$ around 0.5-0.6 
GPa will enhance elastic softening, consistent with the experimental facts in $C_{\rm T}$. The absence of enhanced 
$|T_{\rm Q}|$ and  $E_{\rm JT}$ in $C_{\rm L}$ within the present experimental accuracy remains an open question. Besides 
the hardness of $C_{\rm L}$, which prevents a precise analysis of elastic softening, the difference between $C_{\rm T}$ 
and $C_{\rm L}$ may also suggest that the softening is due to one of the possible quadrupole modes, $\Gamma _3$ or $\Gamma 
_5$, but not both.  

\subsection{Increase of $T_{\rm MI}$ with large Gr\"{u}neisen parameter} 
\begin{figure}[htp]
\includegraphics[width=0.65\linewidth]{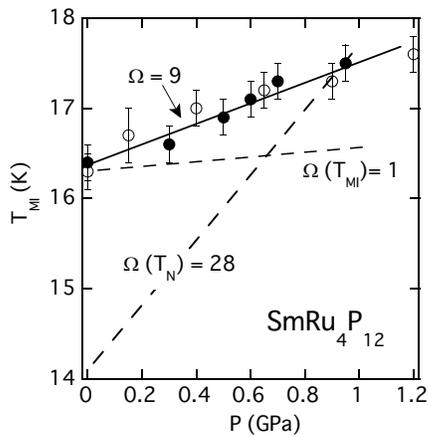}
\caption{$T_{\rm MI}$ as a function of pressure. The solid and open circles denote $T_{\rm MI}$ observed in $C_{\rm T}$ 
and $C_{\rm L}$, respectively.  The solid line is a guide for eyes and represents a Gr\"{u}neisen parameter of $\Omega$ = 
9. The broken lines represent two supposed pressure dependences for $T_{\rm N}$ ($\Omega$ = 28)  and $T_{\rm MI}$ 
($\Omega$ = 1), respectively. 
\label{TmiP.eps}}
\end{figure}

Our results show that $T_{\rm MI}$ increases with pressure (Fig.~\ref{TmiP.eps}), consistent with observations of 
electrical resistance \cite{miyake}. It is also in agreement with thermal expansion measurements where lattice shrinkage 
is observed below $T_{\rm MI}$ \cite{sekine4}.
The monotonic increase of $T_{\rm MI}$ with pressure contrasts strikingly with the unusual changes in $T_{\rm Q}$ and 
$E_{\rm JT}$, suggesting again that the MI transition is not a quadrupolar ordering. Comparatively, no obvious pressure 
dependence \cite{sullow} or a very slight increase \cite{uwatoko} was observed for the $T_{\rm AFQ}$ in the typical AFQ 
compound CeB$_6$.  
Formula~(\ref{eq:4}) defines the Gr\"{u}neisen parameter as volume derivative of the phase transition temperature.
\begin{eqnarray}
\Omega &=& -\frac{\ln T_{\rm MI}}{\ln V} = \frac{C_B}{T_{\rm MI}}\frac{\partial T_{\rm MI}}{\partial P}
\label{eq:4}.
\end{eqnarray}
Based on this formula, the Gr\"{u}neisen parameter $\Omega$ for $T_{\rm MI}$ is estimated to be 9 (solid line in 
Fig.~\ref{TmiP.eps}). The pressure dependence of $T_{\rm MI}$ reported in ref.\,\cite{miyake} leads to an $\Omega$ of 7, 
roughly in agreement with ours. It is significant to make a comparison between the $\Omega$$(T_{\rm MI})$ of 
SmRu$_4$P$_{12}$ and PrRu$_4$P$_{12}$. Performing the same calculation using the reported bulk modulus $C_{\rm B}$ = 207 
GPa \cite{shirotani1} and the pressure dependence of $T_{\rm MI}$ \cite{miyake2}, we obtain $\Omega (T_{\rm MI}) \simeq 2$ 
for PrRu$_4$P$_{12}$. The $\Omega$($T_{\rm MI}$) for PrRu$_4$P$_{12}$ is a normal value while that for SmRu$_4$P$_{12}$ is 
large, again reflecting the essential difference between the two MI transitions. 

The considerably large $\Omega$ is reminiscent of the two very different Gr\"{u}neisen parameters estimated for $T_{\rm 
MI}$ and $T_{\rm N}$ \cite{yoshizawa5}. The conventional equation $\Omega = C_{\rm  B}\beta/C$ (here $\beta$ is volume 
thermal expansion coefficient and $C$ specific heat) gives a huge $\Omega$\,=\,28 for $T_{\rm N}$, but a normal 
$\Omega$\,=\,1 for $T_{\rm MI}$.  
Widely different $\Omega$s are expected to produce widely different pressure dependences of $T_{\rm MI}$ and $T_{\rm N}$, 
as indicated by the broken lines in Fig.~\ref{TmiP.eps}. Unfortunately, no magnetic fields were applied in this work, so 
$T_{\rm MI}$ and $T_{\rm N}$ are not separated  and a more detailed $P$-$T$ phase diagram is not available. Noticeably, 
the $\Omega$\,=\,9 estimated in this work is smaller than that of $T_{\rm N}$ ($\Omega$\,=\,28) and larger than that of 
$T_{\rm MI}$ ($\Omega$\,=\,1). This is an evidence that the pressure dependence of $T_{\rm MI}$ derived from our 
measurements depends not only on the probable multipole OP in phase II but also on the magnetic OP in phase III. 
Instead, we think that separating $T_{\rm N}$ and $T_{\rm MI}$ under pressure is much more difficult than at ambient 
pressure due to the enhanced interplay between the two phases.
In addition, the origin of the enormous $\Omega$($T_{\rm N}$) may be approachable by HF behaviors. A large Gr\"{u}neisen 
parameter is usually observed in HF systems \cite{luthi2} reflecting the large volume effect of the quasi $p$-$f$ 
hybridization band. Indeed, Kondo behaviors (e.\,g., the $-\ln T$ dependence of electrical resistivity) are observed in 
SmRu$_4$P$_{12}$ below 50 K \cite{sekine1} and elastic softening at low temperatures (below 3 K) similar to HF behavior is 
also observed \cite{yoshizawa5}. 

A strong interplay between phases II and III suggested by the large $\Omega$ hints at an octupole scenario for $T_{\rm 
MI}$ in SmRu$_4$P$_{12}$. This is clear because an octupole is magnetic, the same as the dipole OP in phase III, while the 
quadrupole is non-magnetic and holds TRS. In fact, a mix between OP of phase II (octupole) and OP of phase III (dipole) is 
proposed to explain the obscurity of the II-III phase transition in ref.\,\cite{yoshizawa2}. Noticeably, enhanced elastic 
softening under pressure indicates that quadrupole moment may be one of the possible OPs in phase II or III under 
hydrostatic pressure.  We infer two possibilities for this enhanced elastic softening. First, quadrupole ordering is 
induced by pressures in phase II other than the primary OP, so $|g_{\Gamma}|$ and thus elastic softening is enhanced by 
quadrupole fluctuation. Second, quadrupole moment is an accompanying or induced OP in phase III besides the magnetic 
dipole. Therefore, if $T_{\rm N}$ is really approaching $T_{\rm MI}$ as anticipated from the Gr\"{u}neisen parameter 
analysis, the second inference will also produce enhanced elastic softening. There are three kinds of multipole ordering 
available in the $\Gamma _{\rm 67}$ ground state \cite{shiina}, i.\,e., magnetic dipole, electric quadrupole, and magnetic 
octupole. Of these, octupole ordering may inherently induce quadrupole ordering ingredients, as in NpO$_2$ \cite{paixao}, 
or may induce quadrupole ordering by the mode-mixing effect \cite{yoshizawa2, kuramoto}, or by other mechanisms. Thus, 
induced quadrupole OP in phase II or III supports an octupole scenario for the MI transition.  
On the other hand, the largest elastic softening at around 0.5-0.6 GPa means a kind of criticality between the two 
transitions at this pressure, considering the very different Gr\"{u}neisen parameters.  In fact, the virtual temperature 
dependences of $T_{\rm MI}$ and $T_{\rm N}$ based on the Gr\"{u}neisen parameters (broken lines in Fig.~\ref{TmiP.eps}) 
cross at the same pressure. 
When increasing the pressure, the system presumably moves away from the criticality state and consequently the 
quadrupole-strain coupling and quadrupole fluctuation weakens again. This argument needs to be checked theoretically.  
Here it should be noted that we did not observe clear signs that $T_{\rm N}$ exceeds $T_{\rm MI}$ under pressure. So the 
trends of the two phase transition temperatures at pressures higher than the criticality remain an open question.

\section{Summary}
To summarize, we measured the longitudinal elastic constant $C_{\rm L}$ and transverse elastic constant $C_{\rm T}$ for 
polycrystalline SmRu$_{4}$P$_{12}$ under hydrostatic pressure. Significant elastic softening is induced by hydrostatic 
pressure, especially in the transverse $C_{\rm T}$. The results give evidence of degeneracy of the CEF ground state with 
respect to quadrupole moments.
The enhanced elastic softening indicates that quadrupole moment may be one of the possible OPs in phase II or phase III 
under pressure.  
The MI transition is found to increase by applying pressure with a large Gr\"{u}neisen parameter of 9.
We also found that elastic softening above $T_{\rm MI}$ has a strange pressure dependence. The greatest elastic softening 
is observed at around $P$\,=\,0.5-0.6\,Gpa, where the Jahn-Teller energy $E_{\rm JT}$ shows the largest values and 
$|T_{\rm Q}|$ decreases.   
This strangeness may be related to the enormous Gr\"{u}neisen parameter of $T_{\rm N}$ in contrast to that of $T_{\rm 
MI}$, and suggests a strong interplay between phase II and III. 
Our results prefer the octupole moment as the primary OP in phase II.
Experiments in magnetic fields in addition to hydrostatic pressures are currently in progress, in order to separate the 
two successive transitions and establish a detailed $P$-$H$-$T$ phase diagram.
\begin{acknowledgments}
This work is supported by Grant-in-Aid for Scientific Research Priority Area, Skutterudite (no. 15072202) of the Ministry 
of Education, Culture, Sports, Science and Technology of Japan.
\end{acknowledgments}

\end{document}